\documentstyle[a4wide,12pt,epsf]{article}

\newcommand{\GeV}{\mbox{GeV}}
\newcommand{\MeV}{\mbox{MeV}}
\newcommand{\ssc}{{\scriptscriptstyle c}}
\newcommand{\ssb}{{\scriptscriptstyle b}}

\newcommand{\Bc}{B_c}
\newcommand{\bb}{\overline{b}b}
\newcommand{\bc}{\overline{b}c}
\newcommand{\cc}{\overline{c}c}
\newcommand{\mbcresult}{6.386(9)(98)(15)}
\newcommand{\err}[3]{$#1\pm^{#2}_{#3}$}

\begin{document}

\pagestyle{plain}

\begin{flushright}
UTCCP-P-53 \\
GUTPA-98-12-02 \\
Edinburgh 98/23
\end{flushright}
\vskip 20pt

\begin{center}

{\Large \bf  A non-perturbative calculation of the mass of the $B_c$}\\[10pt]
{\bf UKQCD Collaboration} \\[10pt]
{\bf H.P.~Shanahan\footnote{Hugh.Shanahan@physics.org}$^\dagger$, P.~Boyle$^\ddagger$, 
C.T.H.~Davies$^\ddagger$, H.~Newton$^\ast$}\\[10pt]
{\sl $\hbox{}^\dagger$Center for Computational Physics, 
University of Tsukuba,
Tsukuba, Ibaraki 305, 
Japan} \\[10pt]
{\sl $\hbox{}^\ddagger$Dept. of Physics and Astronomy, The Kelvin Building, 
University of Glasgow, Glasgow, G12 8QQ 
Scotland, U.K. }\\[10pt]
{\sl $\hbox{}^\ast$Dept. of Physics and Astronomy, James Clerk Maxwell Building, The King's Buildings, 
University of Edinburgh, Edinburgh, EH9 3JZ, Scotland, U.K.}\\[10pt]
\end{center}
\date{\today}
\begin{abstract}
We present a calculation of the mass of the  $\hbox{}^1S_0$ pseudoscalar 
$\overline{b}c$ ($B_c$) state using a non-perturbative measurement from quenched lattice  
QCD.  We find $M_{Bc} = \mbcresult \; \GeV$ 
where the first error is statistical,
the second systematic due to the quark mass ambiguities and quenching and the third the 
systematic error due to the estimation of  mass of the $\eta_b$.

PACS 12.38.Gc, 12.39.Pn, 14.40.Lb, 14.40.Nd, 14.65.Dw, 14.65.Fy
\end{abstract}


\section{Introduction}
Little is known experimentally
 of the properties of  the $\bc$ spectrum of states,
apart from the detection of the pseudoscalar $\hbox{}^1S_0$, $\bc$ ($\Bc$) state by the 
CDF collaboration \cite{cdf-bc}.
At the very least these states  provide an excellent ``blind'' test for some of the 
techniques used in measuring
non-perturbative properties such as the mass spectrum and decay widths. 
The study of  non-degenerate heavy quarks  provides us with a unique 
probe of QCD.
The weak decays of $\bc$ states provide a plethora of new methods for calculating the less well 
determined  CKM matrix elements, in particular its semi-leptonic decay mode 
\begin{equation}
B_c \rightarrow J/\psi \, l \overline{\nu} \;\; ,
\end{equation}
will provide an accurate measurement of
$|V_{cb}|$ \cite{atlas,vcb}. 

The spectrum of states with the ground state mass subtracted, when calculated via potential 
models \cite{eichten-quigg,fulcher}, for example, is in broad agreement with 
quenched lattice QCD \cite{Andrew,HS,Arifa}. 
Despite concerns that potential models would have large relativistic corrections (as 
a simple centre-of-mass argument implies $\langle v^2/c^2\rangle$ for the charm quark is 
approximately 0.5) it appears that it is still a 
reasonable method for calculating the spectrum.
Nonetheless, a non-perturbative measurement of the spectrum  is important and is being actively
pursued. 
The potential model approach unfortunately does not provide,  {\it a priori}, a
method for determining the ground state mass and hence a number of different 
phenomenological methods have to be used. 
Lattice QCD, on the other hand, can provide an {\it ab initio} method for 
determining its mass.

The use of a non-relativistic 
hamiltonian to describe the behaviour of a bottom quark in quenched lattice QCD
\cite{canonical-NR} 
has been employed very effectively
for the spin-independent sector of the $\overline{b}b$ spectrum \cite{bbar-scaling}.
For relatively coarse lattice spacings, this technique is also applicable  for $\overline{c}c$ states
and indeed two calculations \cite{Andrew,Kim}  of the  $\overline{b}c$ 
spectrum and $m_{B\ssc}$ using non-relativistic $b$ and $c$ quarks have been carried out.
However, as the $\overline{c}c$ spectrum is not reproduced very accurately using this approach
\cite{trottier}
there remains some question as to its accuracy for the $\overline{b}c$ sector. 
Another approach
which will work for lattice spacings where $am_c < 1$ is to use a 
discretised relativistic Lagrangian, 
which, for the lattice spacing used in this letter,
reproduces the charmonium spectrum reasonably well \cite{pab-lat96,pab-lat97}.
Here we take the approach of combining these separate methods for treating 
the $b$ and $c$ quarks. In essence this involves computing Green's functions from a
discretised NRQCD hamiltonian ( NRGF ) with propagators calculated  using a improved 
discretised
QCD Lagrangian devised originally by Sheikholeslami and Wohlert \cite{wohlert}  (  SWP ) 
from the same set of quenched ( zero flavour ) QCD gauge configurations and then 
combining them into two point functions. The damped exponential behaviour of this correlation
function will provide information about the mass of the states we are interested in.

The rest of this letter is organised as follows : in the following section we describe a 
method for computing $m_{B\ssc}$ directly from zero momentum correlation
functions, and demonstrate why this has the smallest systematic error; 
after this  the computational
details of the calculation are outlined,
explaining briefly the details of  NRQCD Hamiltonian used for the bottom
quark and the Lagrangian used for the charm quark;
we then present the numerical results and in the conclusions we compare the results
with other calculations of $m_{B\ssc}$.

\section{Ground state mass definitions}

For large Euclidean times, the two point function of a meson operator constructed 
from either NRGF's or SWP's will take the following 
form

\begin{equation}
\int d^3 x \langle 0 | M({\bf x},t) M^\dagger(0) | 0 \rangle 
\stackrel{\longrightarrow}{\scriptstyle{ t >> 1/m_1}}
A \exp(-am_1 t/a) \;\; ,
\end{equation}
where the integral sign can also indicate a sum over discrete lattice points.
(We have also 
discarded the possible effect of the  finite size of the lattice in the 
time direction for relativistic correlation functions.) 
As demonstrated in 
\cite{mb-nrqcd} and \cite{Andreas}
the variable $am_1$ (where $a$ is the lattice spacing), also 
referred to as ``the pole mass'', has quite different interpretations 
depending on whether the operator $M({\bf x},t)$ is constructed from 
NRGF's or SWP's.
The former is the binding energy of the state with the heavy quark mass subtracted;
the latter satisfies the relationship 
\begin{equation}
am^{SWP}_1 = aM + ac(am_Q,a,V) + ad(a,V) \; \; ,
\end{equation}
where $M$ is the mass of the state, $ad$ are the systematic errors from the finite lattice 
spacing and volume 
and $ac$ is a lattice artefact due to the breaking of Lorentz invariance on  
the lattice
This  parameter vanishes for $am_Q << 1$  and grows to be a large correction as
$am_Q$ approaches 1 and beyond.
Nonetheless, the difference of the pole  masses, for either NRGF's or SWP's
will reproduce the physical
mass difference (assuming other discretisation effects are under control).
Since this process of changing the zero of the quark mass 
is additive, one therefore expects 
\begin{equation}
m_{1B\ssc} - 1/2( m^{NRGF}_{1\eta\ssb} + m^{SWP}_{1\eta\ssc}) 
= M_{B{\scriptscriptstyle c}} - 
1/2 ( M_{\eta{\scriptscriptstyle b}} + M_{\eta{\scriptscriptstyle c}} ) \; ,
\label{mass-diff-eqn}
\end{equation}
where the RHS is the physical mass difference and $m_{1B\ssc}$ is computed using   correlation 
functions with the composition used in this letter.
As one can easily construct a $B$ along the same lines one can also state that 
\begin{equation}
m_{1B\ssc} - ( m_{1B} + m^{SWP}_{1D}) 
= M_{B{\scriptscriptstyle c}} - 
 ( M_{B} + M_{D} ) \; .
\label{mass-diff-eqn-other}
\end{equation}
It is  therefore possible
derive an expression for $M_{B{\scriptscriptstyle c}}$ simply from the
pole masses without resorting to using other possible mass definitions. 

The mass of the $\eta_b$ has yet to be determined, and has to be   estimated 
from the $\Upsilon$ and a potential model calculation of the hyperfine 
splitting.
A larger error is that resulting from the quenched approximation.
A first estimate of the effect of
 quenching is from the variation of the effective lattice spacing between observables having 
different ``typical'' momentum scales. 
In this light we estimate this from the 
difference between the masses determined from equations \ref{mass-diff-eqn} 
and \ref{mass-diff-eqn-other}.

A more traditional approach  to calculate the mass of the $B_c$ is to determine 
the ``kinetic'' mass, (often labelled $am_2$)  which is derived from the dispersion 
relationship of pole masses at different finite momenta. This however has the 
drawback that it has a much larger statistical error than the pole mass.
In particular, while the kinetic mass is proportional to the inverse of a difference
(i.e. the zero and smallest non-zero momentum pole masses), 
the mass derived from equations (\ref{mass-diff-eqn}) and (\ref{mass-diff-eqn-other}) 
are directly proportional to it.


\section{Computational Details}

The calculation was carried out in quenched lattice QCD using 
the Wilson gluon action at 
$\beta=6.2$ on a $24^3 \times 48$ lattice.
Technical details on
how these configurations were calculated and the exact form of the gauge 
action can be found in 
\cite{ukqcd-glue}.
In terms of the string tension resulting from the static quark potential,
the lattice spacing is approximately 0.07 fm, resulting in a spatial 
side length of 1.68 fm.

The non-relativistic Hamiltonian used was corrected to ${\cal{O}}(mv^4)$.
The coefficients in the Hamiltonian were tree-level mean-field estimates.
The mean field improvement coefficient 
is calculated from the plaquette. 
Further details on the Hamiltonian used are described in 
\cite{Andrew}.
The bare mass parameter $aM_b^0$ for the $\bc$ states was chosen
to be $1.22$. 
From \cite{bbar-scaling} this choice of the bare  quark mass parameter 
is consistent,
within quenching errors, with the b-quark mass when defined by the 
kinetic mass of the $\Upsilon$, 
and the scale determined from the $\Upsilon'-\Upsilon$ splitting.
As the splitting of equation (\ref{mass-diff-eqn}) should be 
most sensitive to a variation in the  $b$-quark mass, we use 
this scale to also fix the charm quark mass. In the  case of the $B$-like states we chose
a range of values for $aM_b^0$ from 1.1 to 1.3. A linear interpolation is carried out to determine 
the pole mass for the above bare parameter.

The relativistic Lagrangian used  was the 
Sheikholeslami-Wohlert action \cite{wohlert}.
For the $D$, $\cc$ and $\bc$ states,
the value of the improvement coefficient, $c_{SW}$, in this action 
 was determined by the tadpole improvement procedure 
\cite{mackenzie-tadpole}.
For the $B$ states, a slightly different choice of $c_{SW}$ was used, determined 
by imposing PCAC for light quark masses \cite{alpha-whatever}. For this very fine lattice
spacing, the masses derived from these different prescriptions for $c_{SW}$ are 
indistinguishable \cite{par-thesis}.
For the $B$ and $D$,
a range of light quark
masses were used, and then extrapolated to the chiral limit.

In order to determine the charm quark mass we use  the kinetic mass
of the $D_s$
\cite{pab-lat96,pab-lat97} (requoted in table \ref{kin-mass} for clarity).
This is obtained from correlation functions with one
SWP with either $\kappa = 0.126$ or $0.132$ and one SWP with a $\kappa$
corresponding to the strange quark mass. 
It can be demonstrated that the kinetic masses of these correlation functions  differ
from the pole masses by only a few percent, so ambiguities in the definition of the charm quark 
mass are minimised. In particular we find that the charm quark mass is close to a bare quark mass
equivalent to $\kappa=0.126$, so the interpolation is relatively mild.

In the case of the $\bc$-like  states,
Coulomb gauge fixed correlation functions were computed using sources
optimised to select the ground $\hbox{}^1S_0$ state and its first radial
excitation. The typical radius for these functions  lay in the range
of 3 to 5. 
For $\kappa=0.126$ the correlation functions were computed with 201 configurations
and for $\kappa=0.132$ 140 configurations were used. 
As a Green's function calculated using the NRQCD Hamiltonian
only contains upper spin components, at $\kappa=0.126$
the gauge field was also time reversed so as 
to obtain correlation functions using the lower spin components of the
relativistic propagator and improve the statistics.

Correlation functions for other states in the spectrum, that is the $\hbox{}^3S_1$,
$\hbox{}^1P_1$ and $\hbox{}^3P_{(2,1,0)}$ were also calculated.
The absolute binding energies of the $\hbox{}^1P_1$ and  $\hbox{}^3P_{1,0}$
 were 
calculated while the $\hbox{}^3S_1$ and $\hbox{}^3P_{2}$ energies
 were determined
from single exponential ratio fits relative to the $\hbox{}^1S_0$ and
$\hbox{}^3P_{0}$ respectively.

For the $B$-like states, gauge invariant 
correlation functions were computed with 68 configurations. 
Smeared sources were computed using Jacobi smearing \cite{Jacobi}.
 
The fit criteria for obtaining the pole masses of $\bc$ states were the 
same as those as used
in \cite{ukqcd-tadpole}, using multi-exponential fits.
The $B$-like states were fitted to a single exponential.
 The gauge configurations and SWP's were calculated on a
Cray T3D at the Edinburgh Parallel Computing Centre (EPCC) while the 
Coulomb gauge fixing, NRGF's and correlation functions were computed on a Cray
J90 at the same site.

\section{Results}

\subsection{The ground state mass}
The pole masses calculated for the various heavy-heavy states are listed in 
in table \ref{pole-mass-1}, as well the pole masses for charmonia listed 
in \cite{pab-lat96,pab-lat97}. 
The ``pole mass'' for the $\Upsilon$ obtained from degenerate NRGF's is
listed in table \ref{pole-mass-1} \cite{bbar-scaling}.
The chirally extrapolated results for $D$ and $B$ are quoted in 
tables \ref{pole-mass-2} and \ref{pole-mass-3}.

From \cite{bbar-scaling} the choice of the bare  quark mass parameter of 
1.22 is consistent,
within quenching errors, with the b-quark mass.
On the other hand, the
choice of $\kappa=0.126$ for the relativistic quark is not quite consistent 
with the charm
quark mass.  
By performing a slight linear extrapolation  
on the mass difference defined in (\ref{mass-diff-eqn}) using the kinetic mass, 
we find 
\begin{equation}
am_{1B\ssc} - \frac{1}{2} ( am_{1\eta\ssc} + am_{1\eta\ssb} ) = 0.0557(13) \;\; ,
\label{mass-diff-hh-result}
\end{equation}
and
\begin{equation}
am_{1B\ssc} - ( am_{1B} + am_{1D}) = -0.2480(74)  \;\; .
\label{mass-diff-hl-result}
\end{equation}

As we can see, these two differences vary by a large degree, having a different sign from each other.
When converted into a dimensionful scale, they vary by approximately $1 \; \GeV$.
In the case of equation (\ref{mass-diff-eqn}), the choice of scale is reasonably 
clear;
in order to be consistent in the determination of the quark masses, one should employ the 
scale defined via the $\Upsilon' - \Upsilon$ splitting,
 which for this choice of action
and parameters is $a^{-1}= 3.52(14) \; \GeV$ \cite{bbar-scaling}.
In the case of equation (\ref{mass-diff-eqn-other}), the choice is less clear.
For the  heavy-light masses one would normally use a different definition of the lattice spacing, for
example that defined from the string tension (ideally one would use the $2S-1S$ or $1P-1S$ split
for the appropriate heavy-light combination, but the statistical error is  too large in this case). 
Nonetheless the above
argument for the heavy-heavy difference indicates that the lattice spacing from the 
$\Upsilon' - \Upsilon$ splitting should be used here as well.
 The resulting estimate for $M_{B\ssc}$ from 
equation (\ref{mass-diff-hl-result}) varies by approximately
$200 \, \MeV$ solely by the choice of this ambiguity in the  lattice  spacing. 
The result for the mass from equation (\ref{mass-diff-hh-result}) lies at the centre of this interval.
We therefore take the mass determined from equation (\ref{mass-diff-hh-result}) as our best estimate
of the central value and take the variation with respect to the scale for equation 
(\ref{mass-diff-hl-result}) as a conservative
 estimate of the systematic error due to quenching and quark mass
ambiguities.

We also require an estimate for the hyperfine splitting 
for $\Upsilon - \eta_b$ as experimental data for   
$M_{\eta{\scriptscriptstyle b}}$ is not yet available.
Quenched lattice studies of the charmonium spectrum have always underestimated
the hyperfine splitting (as the central charge from the Coulombic term is smaller 
than usual for 
QCD where $n_f=0$). 
In order to estimate the size of this splitting we use the value of the hyperfine splitting determined 
in \cite{bbar-scaling} in conjunction with the 
the potential model results quoted in 
\cite{eichten-quigg}.
From this we estimate the  $\Upsilon - \eta_b$ splitting 
to be $60(30) \MeV$.
Hence we take $M_{\eta{\scriptscriptstyle b}}= 9.400(30) \; \GeV$.
Using these parameters, one finds :

\begin{equation}
M_{B{\scriptscriptstyle c}} = \mbcresult \;\; \GeV \;\; ,
\end{equation}

where the first error is statistical, the second is the systematic error due to quenching
and the third is the systematic error in the estimation of  $M_{\eta\ssb}$

\subsection{The spectrum}
\begin{figure}
\centerline{\setlength\epsfxsize{90mm}\epsfbox{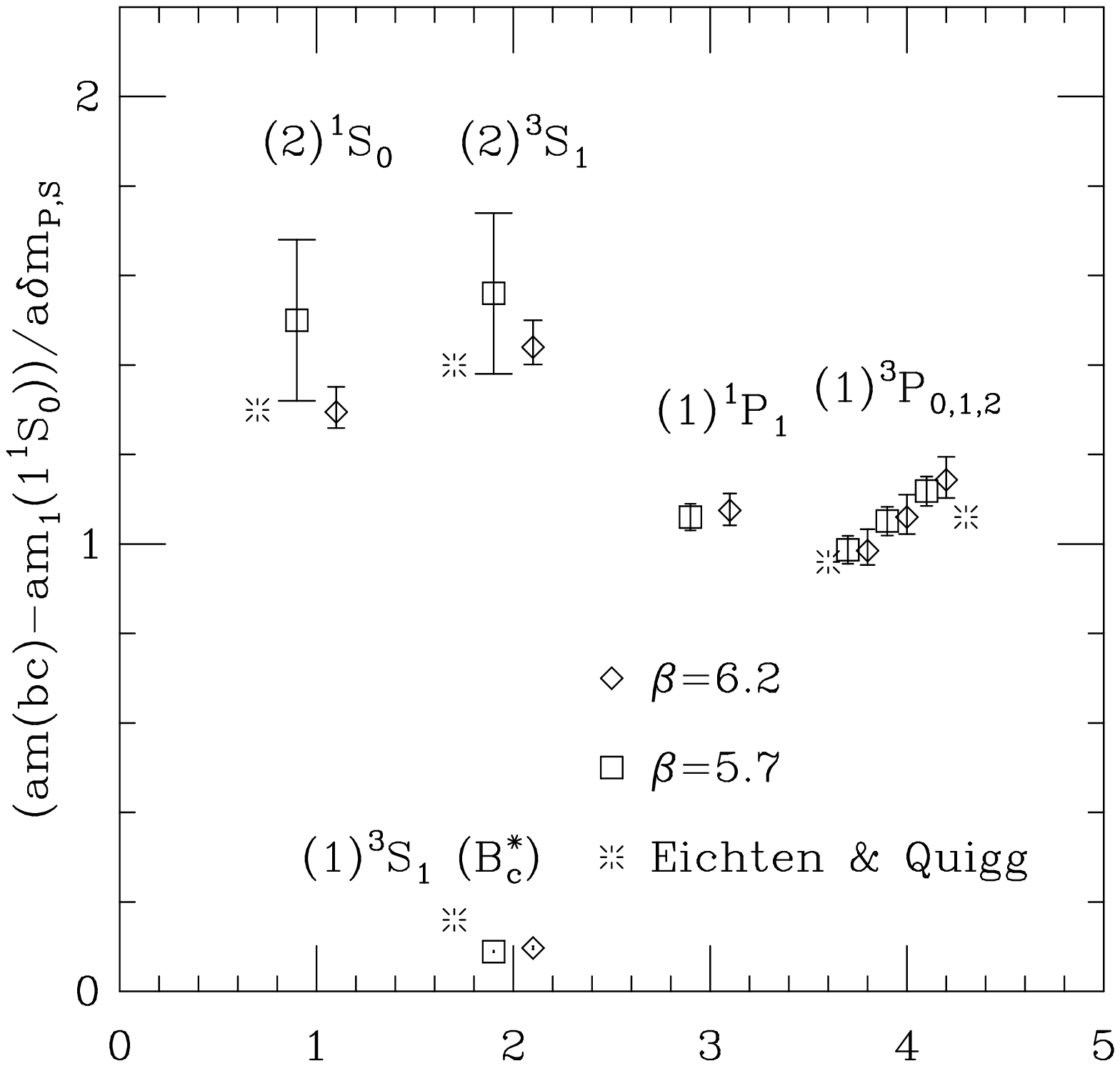}}
 \caption{A comparison of the spectrum of states in this letter with other calculations.
The results denoted by the square and burst  were originally quoted in, respectively in \cite{Andrew}
and \cite{eichten-quigg}. The diamond points represent the results determined in this paper.
To eliminate the scale for the lattice calculations, the spin averaged $1P-1S$ splitting has been
used. The $2E$ and $2T$ labels for the $\hbox{}^3P_2$ states represent
different possible representations in the correlation functions that have
overlap with that state. }
 \label{spectrum-plot}
 \end{figure}

The absolute or relative  binding energies for the higher states of the 
spectrum are listed in 
table \ref{spectrum-table}. 
In figure \ref{spectrum-plot} we plot these results
along with the final lattice results at $\beta=5.7$  quoted in  \cite{Andrew} and the
potential model predictions in \cite{eichten-quigg}, with the scale determined from the spin averaged $1P-1S$ splitting,
which has the smallest statistical errors.
In general these results are in broad agreement with each other, 
although there does appear to be a significant difference in the hyperfine splitting between the lattice results and the 
potential model calculation.

\section{Conclusions}
In this letter we have presented a calculation of the ground state mass of
the $B_c$ and the immediate spectrum of states.
  Although the calculation has been carried in the quenched 
approximation, by using two different possible derivations of the ground state
mass which use a wide range of different possible quark masses, we believe we have 
conservatively estimated its effect. 

The results for the spectrum are, for the most part, in broad agreement with the results
of \cite{eichten-quigg} and \cite{Andrew}.
The ground state mass quoted here  is
consistent with previous lattice calculations \cite{Andrew,Kim,Jones}
in the quenched approximation
calculated using the kinetic mass definition,
using NRGF's for both the charm and bottom quarks on coarser lattices.
QCD sum rule calculations which are listed in \cite{Gershtein} predict that 
$M_{B\ssc}$ lies in the interval of 6.3-6.5 $\GeV$. 
Kwong and Rosner \cite{rosner} surveyed the phenomenological techniques used in determining
$M_{B\ssc}$ and estimated it to lie in the range of 6.194 $\GeV$ to 6.292 $\GeV$.
Most phenomenological estimates for the mass have been taken to lie in the centre of this 
interval \cite{eichten-quigg} although more recently other work by Fulcher \cite{fulcher}
has suggested it to be slightly larger at 
$6.286\stackrel{\scriptstyle{+}}{\scriptstyle{-}}\stackrel{\scriptstyle{15}}{\scriptstyle{6}} 
\, \GeV$.
Nonetheless, the mass we have determined   is certainly 
consistent with the present estimate of the mass from the CDF collaboration
\cite{cdf-bc} of  $M_{B{\scriptscriptstyle c}} = 6.40 \pm 0.39 (stat.) \pm 0.13 (sys.) \GeV$.

Despite the systematic error due to the hyperfine splitting in $\bb$ sector, the largest 
error is the effect of quenching and quark mass ambiguities. 
Eliminating this approximation by the use 
of {\it dynamical}
configurations, where sea quark effects are included 
appears to be the most important step in reducing the error on this mass.
Likewise, a more complete understanding of how the differences of equations (\ref{mass-diff-eqn}) and
(\ref{mass-diff-eqn-other})
behave as a function of the
heavy quark masses also needs to be carried out.
Such configurations (albeit on coarser lattices than used here) are now available
and a measurement of the spin-averaged $B_c$ would be the next course of action.

{\bf Acknowledgements}

\noindent
This work was supported by PPARC and the JSPS Research for the Future
Programme.  
HPS would also like to thank the University of Tsukuba for their hospitality while this
letter was being written. 
The authors are grateful to
the staff at the EPCC for their technical support and Sara Collins, Arifa
Ali Khan and other members of the UKQCD collaboration for useful discussions.
The authors would also like to thank Shoji Hashimoto and Ruedi Burkhalter for
helpful comments and Andreas Kronfeld for pointing out (twice) 
equation (\ref{mass-diff-eqn}).  

\vskip1truecm
\noindent
This paper is dedicated in memory of Andrew Lidsey, one of the 
early researchers of $\bc$ states on the lattice, who died sometime before
this paper was finished. We miss him.

\newpage

\newpage
%
%
\newpage
%

 \begin{table}[t]
\begin{center}
 \begin{tabular}{|c|c|c|c|c|} 
\hline
$\kappa$ & $aM_b$ &  $am_1(\cc)$ & $am_1(\bc)$ & $am_1(\bb)$ \\ \hline
0.126 & 1.22 & 1.035(1) & 0.7207(9) & 0.3027(3) \\ \hline
0.132 & 1.22 & 0.661(1) & 0.5532(18) &  \\ \hline
\end{tabular}
 \caption{Pole masses determined from correlation functions composed of SWP's 
or NRGF's, for heavy-heavy states.}
 \label{pole-mass-1}
\end{center}
\end{table}

\begin{table}
\begin{center}
 \begin{tabular}{|c|c|}
\hline
$\kappa$ &  $am_1(D)$     \\ \hline
0.132 &  0.4332(15)  \\ \hline
0.126 &  0.6424(19)  \\ \hline
 \end{tabular}
 \caption{Chirally extrapolated pole  
masses determined for $D$-like states.}
 \label{pole-mass-2}
\end{center}
 \end{table}

\begin{table}
\begin{center}
 \begin{tabular}{|c|c|}
\hline
$aM_b$ & $am_1(B)$    \\ \hline
1.1  & 0.3192(92) \\ \hline
1.2  & 0.3354(96) \\ \hline
1.3  & 0.347(10) \\ \hline
 \end{tabular}
 \caption{Chirally extrapolated pole  
masses determined for $B$-like states.}
 \label{pole-mass-3}
\end{center}
 \end{table}

 \begin{table}
\begin{center}
 \begin{tabular}{|c|c|} 
\hline
$\kappa$ & $a(\Upsilon' - \Upsilon) / am_2(\overline{c}s)$ \\ \hline
0.126 & 0.255(16) \\ \hline
0.132 & 0.389(23) \\ \hline
\end{tabular}
 \caption{Inverse kinetic masses of heavy-strange correlation functions composed SWF's and the
heavy is in the region of charm. The scale determined from the $\Upsilon' - \Upsilon$ splitting is chosen
in order to maintain consistency with the b-quark mass definition. The physical ratio is
$0.283(3)$.}
 \label{kin-mass}
\end{center}
\end{table}

 \begin{table}
\begin{center}
 \begin{tabular}{|c|c|c|c|} 
\hline
 & $\kappa=0.126$ & $\kappa=0.132$ & $am_{\bc}(m_2(D_s))$ \\ \hline
$aE(\hbox{}^3S_1) - aE(\hbox{}^1S_0)$ & \err{0.0127}{3}{3} &  \err{0.0145}{4}{5}  & \err{0.0131}{3}{3} 
\\ \hline
$aE((2)\hbox{}^1S_0)$ & \err{0.886}{4}{3} & \err{0.756}{11}{7} & \err{0.854}{6}{4}\\ \hline
$aE((2)\hbox{}^3S_1)$ & \err{0.906}{4}{3}  & \err{0.775}{9}{8} & \err{0.876}{5}{4} \\ \hline
$aE(\hbox{}^1P_1)$ & \err{0.864}{5}{3} & \err{0.701}{6}{6} & \err{0.822}{5}{4} \\ \hline
$aE(\hbox{}^3P_0)$ & \err{0.852}{4}{3} & \err{0.691}{4}{6} & \err{0.814}{4}{4}\\ \hline
$aE(\hbox{}^3P_{2T}) - aE(\hbox{}^3P_0)$ & \err{0.021}{2}{2}  & \err{0.018}{5}{3}  & \err{0.020}{3}{2}
\\ \hline
$aE(\hbox{}^3P_{2E}) - aE(\hbox{}^3P_0)$ & \err{0.021}{2}{2}  & \err{0.027}{3}{2} & \err{0.022}{2}{2} 
\\ \hline
$aE(\hbox{}^3P_1) $ & \err{0.861}{5}{4} & \err{0.703}{6}{5}  & \err{0.822}{5}{4} \\ \hline
\end{tabular}
 \caption{Spectrum of states for bare quark masses and the interpolated estimate at the charm mass. }
 \label{spectrum-table}
\end{center}
\end{table}

\end{document}